\documentclass[sigconf,screen]{acmart}

\setcopyright{cc}
\setcctype{by}
\acmDOI{10.1145/3832783.3834549}
\acmYear{2026}
\copyrightyear{2026}
\acmISBN{979-8-4007-2882-2/2026/10}
\acmConference[ASE '26]{Proceedings of the 41st IEEE/ACM International Conference on Automated Software Engineering}{October 12--16, 2026}{Munich, Germany}
\acmBooktitle{Proceedings of the 41st IEEE/ACM International Conference on Automated Software Engineering (ASE '26), October 12--16, 2026, Munich, Germany}
\acmSubmissionID{ase26nier-p62-p}
\received{2026-05-12}
\received[accepted]{2026-07-02}

\usepackage[utf8]{inputenc}
\usepackage[T1]{fontenc}
\usepackage{microtype}
\usepackage{amsfonts}
\usepackage{subcaption}
\usepackage{listings}
\usepackage[inline]{enumitem}

\DeclareUrlCommand{\code}{\urlstyle{tt}}

\definecolor{CodeFrame}{gray}{0.77}
\definecolor{CodeBg}{gray}{0.97}
\definecolor{CodeKeyword}{HTML}{0B63A3}
\definecolor{CodeString}{HTML}{9A3B1C}
\definecolor{CodeComment}{HTML}{4D7C0F}
\definecolor{CodeNumber}{HTML}{7C3AED}
\definecolor{CodeEmph}{HTML}{B91C1C}
\definecolor{CodeLineNo}{gray}{0.45}

\lstdefinestyle{prompt}{
  basicstyle=\ttfamily\small,
  breaklines=true,
  columns=fullflexible,
  frame=single,
  backgroundcolor=\color[RGB]{248,248,248}
}
\lstdefinestyle{code-base}{
  basicstyle=\ttfamily\footnotesize,
  columns=fullflexible,
  keepspaces=true,
  showstringspaces=false,
  upquote=true,
  tabsize=2,
  breaklines=true,
  breakatwhitespace=true,
  frame=single,
  framerule=0.5pt,
  rulecolor=\color{CodeFrame},
  backgroundcolor=\color{CodeBg},
  framesep=1pt,
  aboveskip=0.1em,
  belowskip=0.1em,
  xleftmargin=0pt,
  xrightmargin=0pt,
}

\lstdefinestyle{code-color}{
  style=code-base,
  language=Java,
  keywordstyle=\bfseries\color{CodeKeyword},
  commentstyle=\color{CodeComment},
  stringstyle=\color{CodeString},
  numberstyle=\scriptsize\color{CodeLineNo}
}

\lstdefinestyle{code-print}{
  style=code-base,
  keywordstyle=\bfseries,
  commentstyle=\itshape,
  stringstyle=,
  literate={}
}

\begin{document}

\title{Can Formal Specifications Be Synthesized from Tests Alone?}

\author{Tianhai Liu}
\orcid{0000-0001-5881-1920}
\affiliation{%
  \institution{Karlsruhe Institute of Technology}
  \city{Karlsruhe}
  \country{Germany}
}
\email{tianhai.liu@kit.edu}

\author{Maximilian Müller}
\orcid{0009-0002-5718-8633}
\affiliation{%
  \institution{Karlsruhe Institute of Technology}
  \city{Karlsruhe}
  \country{Germany}
}
\email{maximilian.mueller7@student.kit.edu}

\author{Tobias Hey}
\orcid{0000-0003-0381-1020}
\affiliation{%
  \institution{Karlsruhe Institute of Technology}
  \city{Karlsruhe}
  \country{Germany}
}
\email{hey@kit.edu}

\author{Vitus Lüntzel}
\orcid{0009-0006-5739-0486}
\affiliation{%
  \institution{Karlsruhe Institute of Technology}
  \city{Karlsruhe}
  \country{Germany}
}
\email{luentzel@kit.edu}

\author{Muhammad Minhas}
\orcid{0000-0002-1733-4539}
\affiliation{%
  \institution{Karlsruhe Institute of Technology}
  \city{Karlsruhe}
  \country{Germany}
}
\email{minhas@kit.edu}

\author{Anne Koziolek}
\orcid{0000-0002-1593-3394}
\affiliation{%
  \institution{Karlsruhe Institute of Technology}
  \city{Karlsruhe}
  \country{Germany}
}
\email{koziolek@kit.edu}

\author{Bernhard Beckert}
\orcid{0000-0002-9672-3291}
\affiliation{%
  \institution{Karlsruhe Institute of Technology}
  \city{Karlsruhe}
  \country{Germany}
}
\email{beckert@kit.edu}

\renewcommand{\shortauthors}{Tianhai Liu et al.}

\begin{abstract}
Formal specifications offer strong guarantees, but remain costly to write manually.
Recent LLM-based approaches automate this by inferring specifications from source code, yet their reliance on white-box access poses barriers to industrial adoption due to intellectual property risks and deployment costs.
Our approach uses LLMs to infer candidate specifications solely from test code and dynamic execution traces: the LLM observes only the program interface, selected inputs, and corresponding outputs or state changes, while the implementation internals remain hidden.
Candidate specifications are validated locally using bounded model checking, with feedback guiding iterative refinement. Initial results on the SpecGenBench benchmark suggest that tests can guide LLMs towards meaningful Java Modeling Language specifications, while also highlighting checker compatibility and diagnostic feedback as key challenges for reliable refinement.
\end{abstract}



\maketitle

\section{Introduction}

Unit test suites are the backbone of industrial regression testing, yet they inherently capture only \textit{concrete} executions. 
As systems evolve, this granularity gap allows \textit{silent semantic regressions}
to remain undetected.
While formal specifications (e.g., pre- and postconditions) could bridge this gap by enabling broader behavioral reasoning, they are rarely available in practice due to the manual effort and expert knowledge required to construct them.

Recent efforts to automate specification synthesis have predominantly leveraged Large Language Models (LLMs) in a \textit{white-box} setting, inferring properties directly from source code~\cite{Granberry2025,Ma2025,Puccetti2021,Wen2024,chakraborty_ranking_2023,wu_llm_2024,le-cong_can_2025}. 
While effective, this approach may face practical barriers in industrial contexts. 
Our motivation stems from observations of long-lived industrial systems, such as those maintained by small and medium-sized enterprises (SMEs), where proprietary technology is central to their market differentiation. 
For such systems, maintaining test cases as the software evolves is increasingly challenging, and formal specifications could provide a more durable basis for capturing and checking behavioral expectations.
However, adopting white-box LLM inference must be balanced against the need to protect confidential implementation knowledge and preserve long-term competitiveness.
Moreover, source code may be unavailable to the specification-inference process altogether: security policies may restrict source code access to isolated environments, third-party components or remote services may expose only APIs, and obsolete or legacy systems may survive as binaries after build environments or original source repositories have been lost.
Exposing implementations to external LLMs may be unacceptable, even under contractual safeguards, while maintaining local, state-of-the-art infrastructure is often impractical for SMEs.

These constraints lead us to the research question:
\textit{Can formal specifications be synthesized from tests and their execution observations?}
We assume that such long-lived system implementations are trusted as the reference behavior.
As a first step, we study whether the behavioral knowledge encoded in test code and execution traces can guide LLMs towards meaningful formal specifications without exposing implementation code to the LLM.

In our approach, the program is treated as a black box from the external LLM's perspective: the model is given the program interface, selected inputs, and observed return values or state changes, but not the implementation internals.
Test code is provided to the model, enabling it to exploit oracle assertions, method-call context, and other test-level cues during specification inference.
To complement this static test-side information, executions of the selected tests are encoded as structured observation traces that capture program states before and after the method call under test, including state changes that may not be explicit in the test code. 
The resulting candidates are then iteratively validated and refined in a counterexample-guided loop using bounded model checking (BMC), where the program is \textit{locally} analyzed in a \textit{white-box} manner.

This work presents an initial investigation into the feasibility of LLM-based formal specification synthesis from tests alone. 
Our \textbf{contributions} are:
\begin{enumerate*}
    \item An exploration of the feasibility of decoupling specification inference from source code access, utilizing only test artifacts and execution traces,
    \item a proposed approach that combines oracle information from test code with behavioral observations from test execution, validating candidates via BMC,
    \item a preliminary evaluation on the SpecGenBench benchmark.
\end{enumerate*}

\section{Motivating Example}
\label{sec:example}

\begin{figure*}[t]
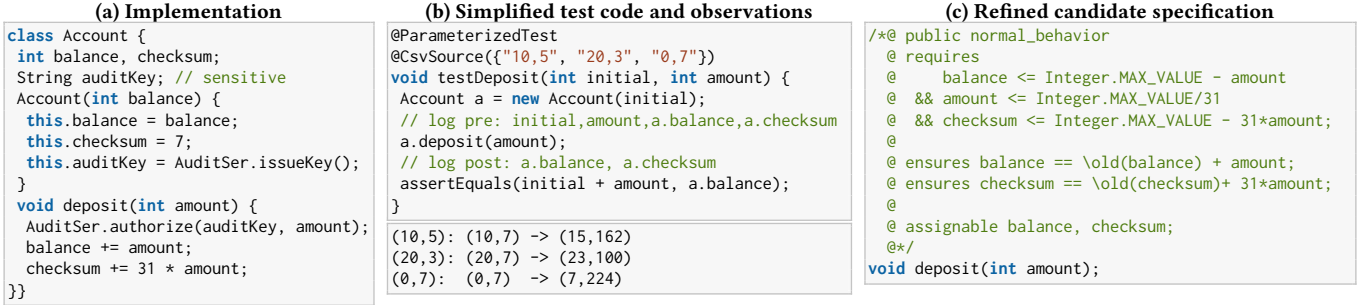

\centering
\tiny

\begin{minipage}[t]{0.27\textwidth}
\subcaption{Implementation}
\label{fig:example-impl}
\begin{lstlisting}[style=code-color]
class Account {
 int balance, checksum;
 String auditKey; // sensitive
 Account(int balance) {
  this.balance = balance;
  this.checksum = 7;
  this.auditKey = AuditSer.issueKey();
 }
 void deposit(int amount) {
  AuditSer.authorize(auditKey, amount);
  balance += amount;
  checksum += 31 * amount;
}}
\end{lstlisting}
\end{minipage}
\hfill
\begin{minipage}[t]{0.34\textwidth}
\subcaption{Simplified test code and observations}
\label{fig:example-test-obs}
\begin{lstlisting}[style=code-color]
@ParameterizedTest
@CsvSource({"10,5", "20,3", "0,7"})
void testDeposit(int initial, int amount) {
 Account a = new Account(initial);
 // log pre: initial,amount,a.balance,a.checksum
 a.deposit(amount);
 // log post: a.balance, a.checksum
 assertEquals(initial + amount, a.balance);
}
\end{lstlisting}
\begin{lstlisting}[style=code-color]
(10,5): (10,7) -> (15,162)
(20,3): (20,7) -> (23,100)
(0,7):  (0,7)  -> (7,224)
\end{lstlisting}
\end{minipage}
\hfill
\begin{minipage}[t]{0.36\textwidth}
\subcaption{Refined candidate specification}
\label{fig:example-spec}
\begin{lstlisting}[style=code-color]
/*@ public normal_behavior
  @ requires 
  @     balance <= Integer.MAX_VALUE - amount
  @  && amount <= Integer.MAX_VALUE/31
  @  && checksum <= Integer.MAX_VALUE - 31*amount;
  @
  @ ensures balance == \old(balance) + amount;
  @ ensures checksum == \old(checksum)+ 31*amount;
  @
  @ assignable balance, checksum;
  @*/
void deposit(int amount);
\end{lstlisting}
\end{minipage}

\caption{Motivating example: inferring a JML specification from tests.}
\Description[Motivating example: inferring a JML specification from tests.]{This figure shows the motivating example that infers a JML specification from tests. The subfigure on the left-hand side shows a program implementation, the middle shows its test code with simplified execution observations, and the subfigure on the right-hand side shows the inferred candidate specification.}
\label{fig:example}
\end{figure*}

Figure~\ref{fig:example} illustrates the setting for a simplified \texttt{deposit} method.
The implementation updates \texttt{balance} and an internal \texttt{checksum}, while \texttt{auditKey} and calls to \texttt{AuditService} represent confidential security or business logic that must not be disclosed.
The test checks only that the balance increases, whereas the engineer-controlled observations record selected non-sensitive state changes before and after the call, e.g., sensitive fields \texttt{auditKey} are excluded.

Using test code and observations, our approach prompts an LLM to synthesize a candidate Java Modeling Language (JML)~\cite{Leavens2006JML} contract.
In this example, the model produces the candidate contract shown in Figure~\ref{fig:example-spec}\footnote{Results may vary across runs and models; this example uses ChatGPT~5.4.}.
The candidate specification is then checked locally using BMC against the implementation via a consistency check and against the observations via a coverage check (see Section~\ref{sec:validation}).
Both checks pass, indicating that the candidate is consistent with the implementation within the analyzed bounds and admits the recorded observations.
Importantly, the inferred contract can capture behaviors not explicitly asserted by the tests but visible in the observations.
For instance, the tests assert only the updated \texttt{balance}, whereas the observations also expose the change to \texttt{checksum}.
Consequently, if the statement updating \texttt{checksum} is removed, the inferred specification detects the violation, although the original test assertions would not.

\section{Proposed Approach}
\label{sec:approach}

\begin{figure}[tb]
  \centering
  \includegraphics[width=0.8\linewidth]{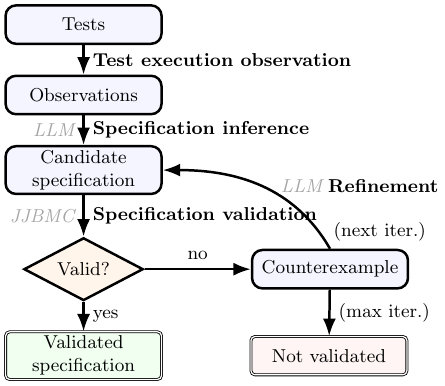}
  \caption{Overview of the specification inference pipeline.}
  \Description[Overview of the specification inference pipeline.]{This figure shows the specification inference pipeline.}
  \label{fig:architecture}
\end{figure}

Our approach infers specifications from test code and execution observations.
As shown in Figure~\ref{fig:architecture}, tests are first executed to produce test execution observations, which make relevant pre- and post-state explicit.
An LLM then uses the tests and observations to synthesize a candidate JML specification.
The candidate is validated locally with BMC against the implementation and the recorded observations.
If validation fails, the counterexample is fed back to the LLM for refinement. 
Otherwise, the candidate is returned as a specification that admits the observed test behaviors and is consistent with the implementation within the analyzed bounds.

\subsection{Test Execution Observation}
\label{sec:observation}

We execute selected unit tests for the method under test to obtain concrete execution traces.
These traces complement test code by making observed input-output and state-transition relations explicit, which can refine specifications beyond assertions alone and provide behavioral cues when interface names are anonymized.
Test selection may be guided by domain expertise, coverage goals, and operational constraints.
For each selected test, engineers specify which objects and fields should be recorded immediately before and after the method call.
This defines a user-controlled projection of the observable program state, which can be implemented through logging or bytecode instrumentation.
In the motivating example (Figure~\ref{fig:example}), we use logging statements for ease of presentation.
Tests that expose sensitive information through their code, inputs, or recorded execution data are excluded.

From each execution trace, we extract a test execution \textit{observation} that makes the behavioral evidence explicit and reusable for prompting, observation-coverage checking, and refinement (Section~\ref{sec:inference}--\ref{sec:validation}).
Each observation corresponds to one method invocation and is represented as
\(obs = \langle args,\; pre,\; post,\; outcome \rangle\).
The \(args\) records the receiver-construction arguments and method-call arguments, including parameters of parameterized tests.
The \(pre\) and \(post\) components record the selected program state immediately before and after the invocation, respectively.
The \(outcome\) component records either normal termination, including the return value, or an exception.
Objects in \(pre\) and \(post\) are represented by an identifier, a class name, and field-value mappings.

\subsection{Specification Inference}
\label{sec:inference}

We employ an LLM to synthesize a method-level JML specification from the method signature, the tests that invoke the method, and the corresponding observations.
The zero-shot prompt follows a two-stage structure.
First, the model infers an initial contract from the test code, using test inputs and assertions as evidence for candidate preconditions, postconditions, exceptional behavior, and assignable locations.
Second, the model refines this contract using the observations, in particular repeated pre/post-state relations and observed return or exception outcomes.
To reduce unsupported generalization, the prompt enforces three rules.
First, clauses should generalize beyond individual tests only when supported by test assertions, repeated observations, or validation feedback.
Second, unsupported domain assumptions should be reported as hypotheses rather than encoded directly as JML clauses.
Third, the final output should contain non-vacuous, syntactically valid JML clauses.

\subsection{Specification Validation}
\label{sec:validation}

Given a candidate specification inferred from test code and observations, we validate it using two complementary checks. The consistency check detects candidate specifications that are contradicted by the implementation within the analyzed bounds. The observation-coverage check ensures that the inferred specification does not exclude any observed test behavior.

We consider three behavior sets:
(i) $B_{code}$, behaviors exhibited by the implementation,
(ii) $B_{test}$, the subset exercised by the selected unit tests, and
(iii) $B_{spec}$, behaviors admitted by the candidate specification, with $B_{test} \subseteq B_{code}$ by construction.
A behavior $b$ corresponds to a pre- and post-state pair of a method execution.

\textbf{Consistency check.}
We treat the implementation as the local behavioral reference, reflecting our target setting of mature industrial software that has evolved over the years and is already supported by regression tests.
A candidate specification should therefore not exclude behaviors that the implementation can exhibit within the bounds of analysis.
We use BMC to check, within a given bound, whether the precondition and program semantics admit an execution that violates the postcondition.
If such an execution exists, BMC reports it as a counterexample \(b \in B_{code} \setminus B_{spec}\), showing that the specification excludes an implementation behavior and is therefore inconsistent.
We convert the counterexample into an additional observation and feed it back to the LLM for refinement.

\textbf{Observation-coverage check.}
The consistency check ensures that the candidate specification admits all implementation behaviors within the analyzed bound, e.g., bounded loop unrollings.
If this bound is sufficient, the observed execution behaviors would already be admitted by the specification.
However, tests may exercise behaviors outside the BMC search space.
We therefore perform an observation-coverage check that directly checks whether each recorded test behavior is admitted by the candidate specification.

We encode observations as JML-annotated harness specifications that fix observed pre-states and require the corresponding post-states.
The target method is interpreted contractually rather than inlined, making the check independent of the BMC loop-unrolling bound.
A counterexample \(b \in B_{test} \setminus B_{spec}\) means that the contract excludes an observed behavior and guides refinement.
For example, an observation \(balance:10\mapsto15\) under \(amount=5\) means that the contract allows a call starting from \(a.balance=10\) with \(amount=5\) and ending in \(a.balance=15\).
A candidate with \texttt{requires amount > 10} rejects this call and therefore fails the coverage check.

A candidate that passes both checks is treated as a validated candidate, not as a complete specification.
The checks rule out two concrete failure modes: contradicting the implementation within the analyzed bounds and excluding recorded test behaviors.
They do not establish that all unobserved behaviors are specified precisely; specification strength is therefore assessed separately.

\section{Preliminary Results}
\label{sec:preliminary-results}

\begin{table}[t]
\centering
\caption{Preliminary results on the 22 SpecGenBench tasks.}
\label{tab:preliminary-results}
\footnotesize
\setlength{\tabcolsep}{0pt}
\renewcommand{\arraystretch}{1}
\begin{tabular}{l@{\hspace{0.5em}}c@{\hspace{0.5em}}l@{\hspace{0.5em}}|@{\hspace{0.5em}}l@{\hspace{-0.5em}}c@{\hspace{-0.1em}}r}
\toprule
\textbf{Task} & \textbf{History} & \textbf{Result} &
\textbf{Task} & \textbf{History} & \textbf{Result} \\
\midrule
\texttt{abs} & \texttt{11} & Success &
\texttt{binarySearch} & \texttt{00} & Error \\
\texttt{biggest} & \texttt{11} & Success &
\texttt{bubbleSortDesc} & \texttt{00} & Error \\
\texttt{calculate} & \texttt{11} & Success &
\texttt{arrcmp} & \texttt{00} & Error \\
\texttt{canWinNim} & \texttt{11} & Success &
\texttt{containsDuplicate} & \texttt{00} & Error \\
\texttt{conjunctOf} & \texttt{11} & Success &
\texttt{convertToFahrenheit} & \texttt{00} & Error \\
\texttt{disjunctOf} & \texttt{11} & Success &
\texttt{convertToKelvin} & \texttt{00} & Error \\
\texttt{divisorGame} & \texttt{11} & Success &
\texttt{convertToTitle} & \texttt{00} & Error \\
\texttt{addLoop} & \texttt{01,11} & Success &
\texttt{copyArray} & \texttt{00} & Error \\
\texttt{checkString} & \texttt{01,11} & Success &
\texttt{digitRoot} & \texttt{00} & Error \\
\texttt{computeOverlapArea} & \texttt{01,11} & Success &
\texttt{bubbleSort} & \texttt{00,00,00} & Fail \\
\texttt{convertTemperature} & \texttt{10,10,11} & Success &
\texttt{changeCase} & \texttt{10,10,10} & Fail \\
\bottomrule
\end{tabular}
\end{table}

We conducted an initial feasibility study on SpecGenBench~\cite{Ma2025}. 
SpecGenBench was originally designed for specification inference from source code, where multiple implementations of the same API provide complementary evidence about the intended behavior. 
In contrast, our approach infers specifications from tests and execution observations without exposing the implementation source code. 
We therefore evaluate our approach at the API level, treating each API as a separate target-specification task.

For each API, we use \textit{gpt-oss-120b}~\cite{openai2026gpt-oss-120bapi} to generate JUnit tests. 
Across the evaluated tasks, these tests achieve an average instruction coverage of 76.9\%, providing a coarse sanity check that the inferred specifications are not based solely on trivial executions. 
We use GPT-5.4~\cite{openai2026gpt54api} to infer and refine candidate JML specifications.
Each candidate is checked with JJBMC~\cite{Beckert2020} using its default loop-unrolling bound of 5 against two criteria: implementation consistency and observation admission. Such bounds are a known source of precision concerns in bounded verification~\cite{Liu2017LoopBounds}.
Failed candidates are refined using validation feedback for up to 3 iterations per task.
We evaluate our prototype on 22 SpecGenBench APIs compatible with JJBMC. 
Their implementations use Booleans, integers, floating-point values, arrays, and loops.
We exclude tasks that require features JJBMC doesn't support, such as strings or reachability-style heap properties. 

Table~\ref{tab:preliminary-results} shows the evaluation results. 
We classify the outcomes as follows: \textsc{Success} means that the final candidate passes both the consistency and observation-coverage checks; \textsc{Error} means that the generated JML is not supported by JJBMC; and \textsc{Fail} means that the refinement iteration bound is reached without obtaining a candidate that passes both checks.
In the history column, each iteration is encoded as a two-bit outcome: the first bit denotes the consistency check, and the second denotes the observation-coverage check; 1 indicates success and 0 indicates failure. 
For example, \(01,11\) means that consistency fails but coverage passes in the first iteration, while both checks pass in the second.

The prototype succeeds on 11 of the 22 tasks. 
Among these, 7 are solved in the first iteration, while 4 require refinement.
The \texttt{convertTemperature} case is representative of the refinement. 
The initial specification expressed the expected conversion results as postcondition $\texttt{result}[0] = \texttt{celsius} + 273.15$, where \texttt{celsius} is the parameter of the tested method. During refinement, JJBMC reported a consistency violation caused by unsafe JML expressions, such as dereferencing \texttt{\textbackslash result} or accessing array elements without guards. 
Using this feedback, the LLM added well-definedness guards \texttt{\textbackslash result != null \&\& \textbackslash result.length >= 1}
\(\Rightarrow\)
\texttt{\textbackslash result[0] == celsius + 273.15}.
Such guards avoid null dereferences and out-of-bounds accesses during JML evaluation. 

The dominant limitation is compatibility with the JML fragment supported by JJBMC. 
Among the 22 tasks, 9 terminate with an \textit{Error} because the generated specifications contain constructs outside the fragment supported by JJBMC, such as \texttt{\textbackslash num\_of}, ghost or model fields, and arithmetic expressions over quantified variables.
Since these constructs are valid in standard JML but cannot be validated by the current toolchain, we treat these cases primarily as checker-compatibility failures rather than semantic failures of the inferred specifications after closer manual investigation.

Checker interaction also explains the two \textit{fail} cases. 
These tasks reach the refinement bound without obtaining a candidate that is both consistent with the implementation and covers the observations. 
In these cases, JJBMC reports failed checks but does not provide an informative counterexample, returning only an assertion failure with an empty concrete example. 
Consequently, the LLM receives little actionable feedback and can only guess how to refine the candidate specification.

The results provide early evidence for the feasibility of inferring JML specifications from test code and observations, while identifying checker-aware specification generation as the main current challenge. 
Future work will therefore provide checker-aware pre-checking and a repair layer, and improve JJBMC.

A threat to validity is the potential leakage of benchmark data. SpecGenBench tasks or similar code fragments may have appeared in LLM pre-training data. 
We plan to mitigate this risk by evaluating the approach on proprietary industrial software whose implementations and intended specifications are not publicly available, and by anonymizing the method and test code in SpecGenBench.

\section{Related Work}
\label{sec:related-work}

Dynamic specification inference derives likely properties from observed executions. 
Daikon~\cite{Ernst1999} pioneered this line by mining invariants from instrumented runs, while later work on specification mining inferred temporal rules, API protocols, behavioral patterns, and algebraic specifications from execution traces and program interactions~\cite{ammonsMiningSpecifications2002,henkelDiscoveringAlgebraicSpecifications2003}.
Subsequent approaches extended the predicate space with quantified properties, implication-style specifications, or evolutionary search~\cite{Wei2011,Meyer2007,Molina2021}.
These techniques are related to our use of observations, but they typically rely on predefined templates, handcrafted predicates, or trace mining rather than exploiting test-code oracle structure and LLM-based generalization.

Specification synthesis approaches such as AutoSpec~\cite{Wen2024}, SpecGen~\cite{Ma2025}, and Allo2JML~\cite{Grunwald2014JML}, as well as works on loop invariant generation~\cite{Beckert2025,Teuber2025,chakraborty_ranking_2023,pirzadaLLMGeneratedInvariantsBounded2024} and class invariant generation~\cite{sunClassInvGenClassInvariant2026}, demonstrate the potential of automated specification synthesis.

Granberry et al.~\cite{Granberry2025} further enrich prompts with generated tests and static analysis results.
In contrast, our work explores a complementary setting in which the LLM receives controlled test code and sanitized observations, while implementation code is excluded from the prompt and used only for local validation.


Test- and oracle-based techniques infer executable assertions or postconditions from generated executions or transformation traces~\cite{Xie2006Orstra}. 
Like our approach, they treat executions as evidence of intended behavior.
However, they usually aim to strengthen test oracles or learn assertions, rather than to synthesize JML method contracts. 
Our approach uses test code structure and heap observations as evidence for LLM-based generalization, followed by local JML validation.

\section{Conclusion and Future Work}
\label{sec:conclusion}

We investigated whether LLMs can synthesize formal specifications from tests alone. Our framework extracts structured execution observations, combines them with test-oracle information to generate candidate JML specifications, and validates them through BMC-based counterexample-guided refinement. This behavior-oriented approach avoids reliance on source-code structure. On 22 SpecGenBench API-level tasks, our prototype automatically generated and validated specifications for 11 using JJBMC.

Test-level evidence provides partial behavioral coverage. To offset missing source-code context, we plan to incorporate public documentation~\cite{Zhong2009}, API specifications~\cite{Pandita2012NL2Spec}, and usage examples~\cite{Raychev2014CodeCompletion} to capture intent and resolve behavior under-specified by tests.

We will also address scalability by summarising large traces into representative behavioral patterns before inference. To mitigate BMC state-space explosion, we plan to combine bounded validation with program slicing~\cite{Weiser1984,Tip1995} and incremental symbolic execution~\cite{Pasareanu2010,Liu2014IncrementalSE}.

\section*{Acknowledgements}
This work was supported by the German Research Foundation (DFG) – SFB 1608 – 501798263.

\balance

\section*{Data Availability Statement}
All benchmark, evaluation data, and prototype artifacts are publicly available at~\cite{artifact}.

\bibliographystyle{ACM-Reference-Format}
\bibliography{references}

\end{document}